# Context-Aware DFM Rule Analysis and Scoring Using Machine Learning

Vikas Tripathi, Valerio Perez, Yongfu Li, Zhao Chuan Lee, I-Lun Tseng, and Jonathan Ong
GLOBALFOUNDRIES, 60 Woodlands Ind. Park D Street 2, 738406, Singapore

**Abstract -** To evaluate the quality of physical layout designs in terms of manufacturability, DFM rule scoring techniques have been widely used in physical design and physical verification phases. However, one major drawback of conventional DFM rule scoring methodologies is that resultant DFM rule scores may not accurate since the scores may not highly correspond to lithography simulation results. For instance, conventional DFM rule scoring methodologies usually use rule-based techniques to compute scores without considering neighboring geometric scenarios of targeted layout shapes. That can lead to inaccurate scoring results since computed DFM rule scores can be either too optimistic or too pessimistic. Therefore, in this paper, we propose a novel approach with the use of machine learning technology to analyze the context of targeted layouts and predict their lithography impacts on manufacturability.

## I. Introduction

Designs for manufacturability (DFM) rules are a set of recommended rules which are aimed at improving physical design quality in terms of manufacturability. These rules can be considered as extra rules in addition to design rule checking (DRC) rules, although part of DFM rules can be optional. A design is considered to have satisfying manufacturability if the design does not violate any DFM rule. At GLOBALFOUNDRIES, we use a DFM rule scoring methodology to evaluate physical design quality in terms of manufacturability. In the scoring methodology, each DFM rule violation is provided with a DFM score, ranging from 0 to 1, based on the severity of the DFM rule violation; scores that are close to zero are considered as severe violations.

As transistor feature sizes continue shrinking, the optical resolution for lithography tools has increased significantly, which makes the role of layout context very critical in printability of any shape and can significantly impact the effective margins required to print them clean. Layout shapes that are not lithography-friendly can cause printability issues although they can be identified by using lithography simulations. Additionally, these layout shapes can detract the printability of neighboring geometries if they are placed in close proximity and thus can be considered as unfriendly layout context. In the scenario where a DFM violation is surrounded by such unfriendly layout context, the severity of such violation will increase and thus the chances of having functional failures will be higher. Conventional DFM rule scoring methodologies do not consider layout context while performing margin analysis and scoring, making it ineffective to identify DFM violations which are more severe due to its unfriendly layout context. This limitation creates misalignment in rule-based analysis and lithography results, making these methodologies unable to produce accurate scoring results. Such limitation also makes them difficult to filter or bin critical DFM violations separately.

Figure 1 illustrates the limitation of conventional DFM rule scoring methodologies. In figure 1(a), the input layout has two metal-via DFM enclosure rule violations, which are labeled as "1" and "2". Both violations violate the same enclosure value. Conventional DFM rule scoring methodologies treat them equally as both have the same enclosure and thus produce the same score for both violations, as shown in figure 1(b). However, lithography simulation results for both violations may not be the same. One violation fails to print the required enclosure whereas the other can be printed successfully, as shown in figure 1(c).

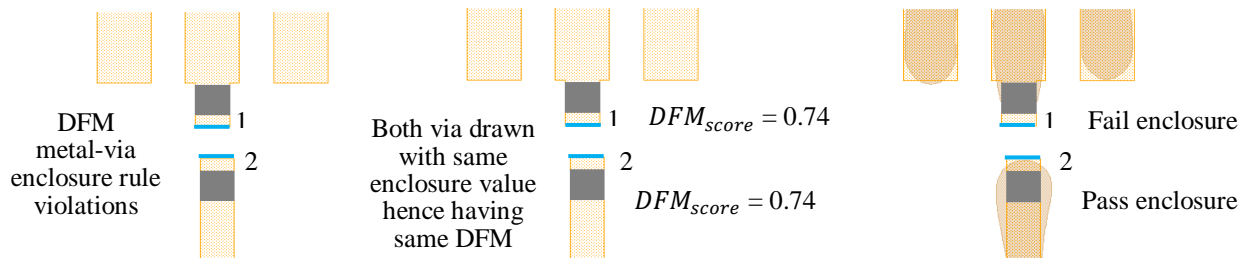

**Figure 1:** (a) Layout of two DFM violations (b) conventional scoring output (c) lithography simulation results.

This mismatch is due to the difference in their layout context, violation "1" has neighboring line-ends and connected jogs; they can affect the printability significantly. The other violation does not have such neighboring shapes and thus has better printability. The limitation of the conventional methodologies creates a need for developing advanced techniques for considering layout context while performing rule-based scoring.

In this paper, we propose a novel methodology for performing context-aware rule checking and scoring. This methodology is based on machine learning so that problematic layout contexts from past lithography simulation results are used as a knowledge base and then use this knowledge base to predict a given layout context for its probability of causing failures. In this paper, we propose a methodology by providing an overflow of the general flow and use neural networks with machine learning techniques to predict lithography impacts for a layout context. We also discuss about the generation of training set data for training neural networks, followed by the integration with rule-based verification and score computation. Finally, conclusions are drawn in Section V.

## II. Method

In this proposed model, Artificial Neural Network (ANN) is applied and trained using supervised machine learning technique. This is achieved using a set of labeled data which has been extracted from past results. The training process of ANN will help to compute weights and biasing for different synapsis present in the neural network, which will be used during the process of DFM rule scoring.

Figure 2 shows the flow diagram of context analysis. In this flow, Data extraction is performed and used in training neural network. This data is prepared using previously reported lithography hotspots with DFM rule violations and overlay them with corresponding layout design. If a DFM violation is overlaid with lithography hotspot then we classify it as a "bad context" and if it does not have any issue in printability then we consider it as a "good context". This results into classification of data in good and bad categories. The good category violations are still lithography friendly while bad category violation are not lithography friendly and considered more severe. Next, layout context specification will be translated in to metric form for each DFM rule violation which will be used as initial input for the neural network.

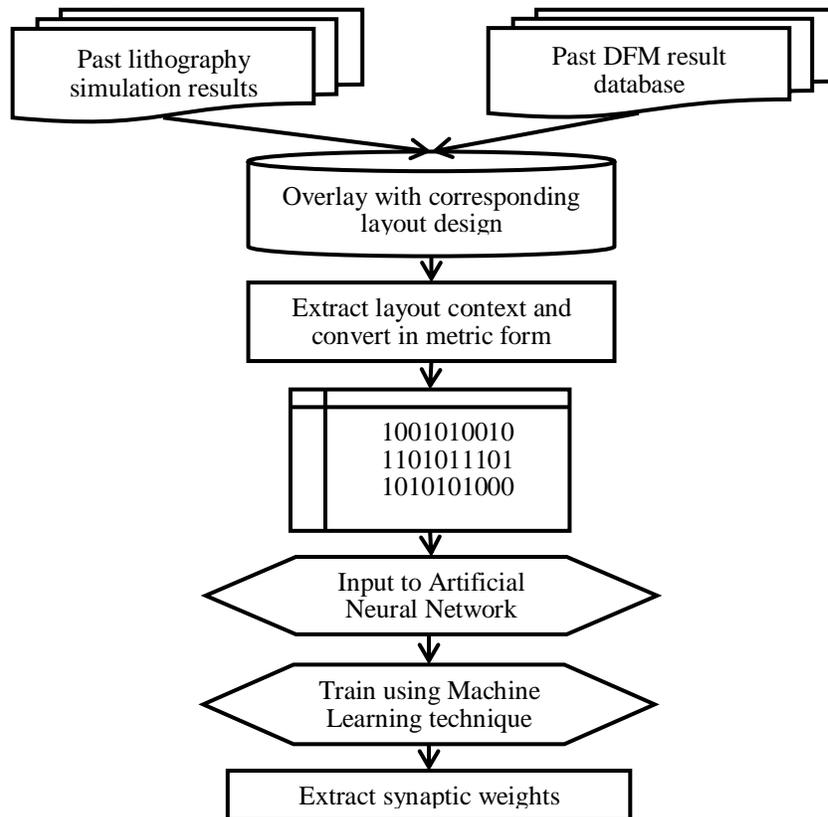

**Figure 2:** Context Analysis flow with Machine Learning

Figure 3 shows two sample DFM enclosure violations edges, highlighted in blue color, along with their layout context. The edges are potential candidate to characterize layout context. The DFM violation 3(a) is also a lithographically hotspot whereas DFM violation 3(b) is clean. To extract the layout context, we divide the context in 8 different regions. Each region is analyzed for layout polygon and vertices count. These values are then used to build a metric, which is then appended with lithography simulation result. If DFM violation overlay with lithography hotspot then "1" is appended to the metric otherwise a "0" is appended.

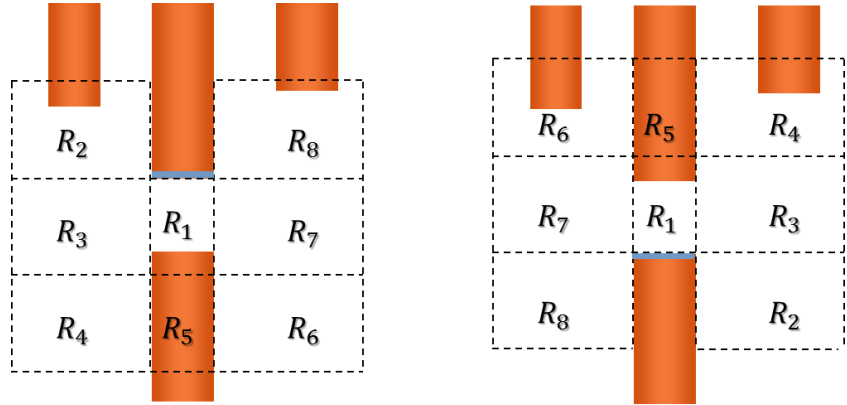

**Figure 3:** Layout context of (a) DFM violation which is also a Lithography hotspot (b) DFM violation but Lithography clean

The tabulation of final representation is shown in table 1. The first row represents context sequence for all the DFM violation. This sequence remains constant for all the violations considering rotation and placements. Similarly entire layout context can be extracted for all the DFM violations available in multiple designs. This generates huge amount of data which is required to build and train our artificial neural network. In the next section, discussion on how to build and train artificial neural network will be followed.

|      | $R_1$ | $R_2$ | $R_3$ | $R_4$ | $R_5$ | $R_6$ | $R_7$ | $R_8$ | Result |
|------|-------|-------|-------|-------|-------|-------|-------|-------|--------|
| 3(a) | 3     | 3     | 0     | 0     | 1     | 0     | 0     | 3     | 1      |
| 3(b) | 3     | 0     | 0     | 3     | 1     | 3     | 0     | 0     | 0      |

**Table 1:** Layout context representation in metric form

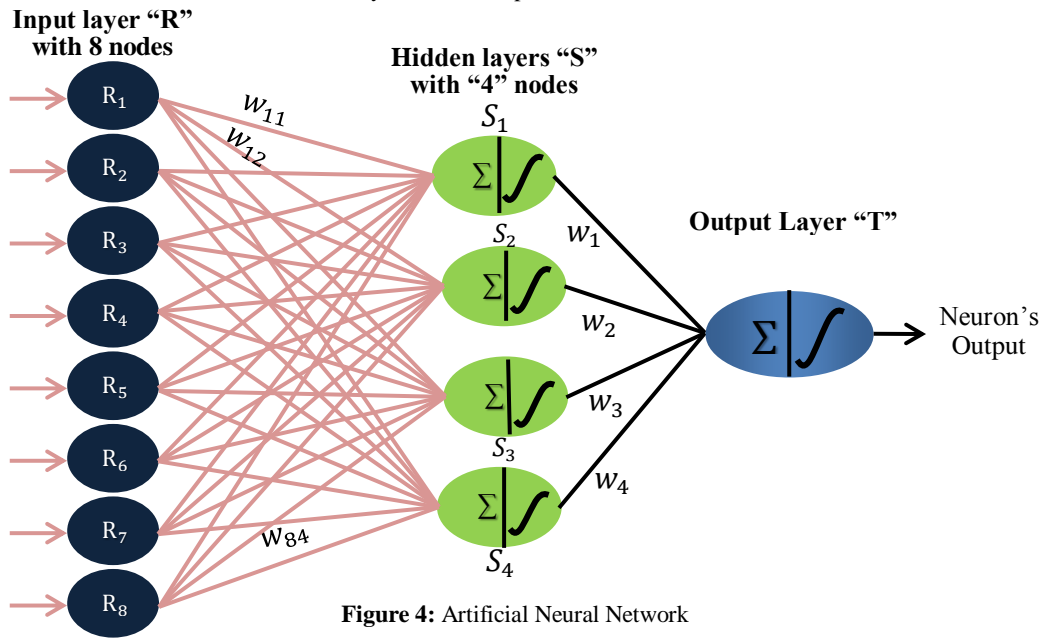

**Figure 4:** Artificial Neural Network

Figure 4 represents the neural network. Eight input nodes (considered as input layer) represents the value of each region extracted from layout context. The outputs of these layers are connected with next intermediate layer in the network using a weighted connection. These weights represent the significance of associated node in neural network. This intermediate layer is also considered as a hidden layer because it has no direct connection with the outside system. These layers are helpful in performing computations and transfer information from the input nodes to the output nodes. Every hidden layer can introduce a non-linearity in the function which can be helpful in order to solve complex nonlinear correlations between input and output. In this approach, only one hidden layer is used to get the optimum results using it. The output of hidden layer nodes are connecting to single node in output layer using another weighted connection. In general the purpose is to keep the network optimized and avoid any redundant nodes or over fitting. This is based on the quantity and quality of input database. A strong data is crucial in strong prediction and perform well even with less neuron in the network. On the other hand if the data is week and noisy then it might require more set of neurons with multiple hidden layers. If there are any changes in the input data set then one can recalibrate the network for optimal results. To begin with this network optimization and training process, known data sets extracted from past sample designs is used. In this case, over multiple data points (known cases) with expected output is applied and further can be used as a training set to predict the outcome for new unknown cases.

|  | Input Set | Neuron Output |
|---|---|---|
| Known Case 1 | 310212011 | 1 |
| Known Case 2 | 221001000 | 0 |
| Known Case 3 | 341010100 | 1 |
| Known Case 4 | 101022101 | 1 |
| ..... | ..... | .. |
| .. | .. | . |
| **New Case** | **22300102** | **?** |

**Table 2:** Example of a training set

To initiate the process, each input is given a random weight which can be a positive or negative number. These weights impact the effectiveness of associated input. Inputs having a large positive or negative weight will have a stronger effect on the neuron's output. The weighted output for each node of input payer is calculated using equation 1. In this equation $s_{j_{in}}$ represents the weighted sum of input layer where $j$ represent the hidden node number, $w_{ij}$ represent the synaptic weight connected between $i^{th}$ input node to $j^{th}$ hidden node and $R_i$ represents the $i^{th}$ input node value. The intermediate nodes sum up all the weighted inputs and use sigmoid function to normalize the summation results. Here sigmoid function is used as to have a normalized output scaling from 0 - 1 as shown in equation 2. Here $s_{j_{out}}$ represent the final value of a $j^{th}$ hidden layer node.

$$s_{j_{in}} = \sum_{i=1}^{8} w_{ij} * R_i \qquad \text{(Equation 1)}$$

$$s_{j_{out}} = \frac{1}{\left(1+e^{-s_{j_{in}}}\right)} \qquad \text{(Equation 2)}$$

Similarly we calculate the value for output node. Similarly, equation 3 represents the weighed sum of hidden layer nodes as $t_{in}$, here $w_j$ represent the synaptic weightage connected between $j^{th}$ hidden layer nodes and output node where $s_{j_{out}}$ defines the $j^{th}$ hidden layer node value. To calculate the final neuron's output another sigmoid function is applied as shown in equation 4 here $t_{out}$ represent the final neuron's output.

$$t_{in} = \sum_{j=1}^{4} s_j w_j \qquad \text{(Equation 3)}$$

$$t_{out} = \frac{1}{(1+e^{-t_{in}})} \qquad \text{(Equation 4)}$$

The main variable for equation 1 and 3 are the synaptic weightages. As we start with some random weights to these synaptic hence the result will be not be accurate. To correct the outcome of this neuron we need to adjust the weights according to their significance, which is done using a Train, Predict and Adjust cycle defined in figure 5.

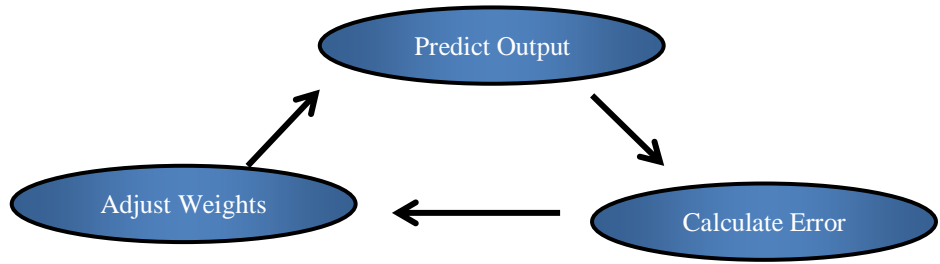

**Figure 5 :** Training Cycle

In this cycle the predicted outcome is used to calculate the error which is the difference between the neuron's output and the desired output from the training set example and adjust the weights in such a way that it reduces the error value. To implement this we have used gradient descent learning rule [1] for updating the weights by calculating the gradient of the loss function. This technique is also called backward propagation of errors, because the error is calculated at the output and distributed back calculating through the network layers. The generalized equation 5 helps in adjusting the synaptic weights which is proportional to the size of the error. If the output is a large positive or negative number, it signifies the neuron is confident on prediction and does not make much adjustment in the output. Similarly, the weights synaptic between input and hidden layer also need to be adjusted. For this we use chain rule. In this the adjusted weight of hidden layer nodes are used to calculate the weights of one layer below the chain. Figure 6 showing the complete learning algorithms.

$$Adjust\_Weights \mathrel{+}= error * input * SigmoidGradent(output) \qquad \textbf{(Equation 5)}$$

```
1. Define ANN architecture (8:4:1) and initializes weights randomly
```
   $w_{ij}$ = random weights for synapsis connected from input to hidden layer
   $w_j$ = random weights for synapsis connected from hidden layer to output neuron
```
2. Define input vector to the neuron
```
   $R_i$ = input layer vector
   $s_{j_{out}}$ = hidden layer vector (using equation 2)
   $t_{out}$ = out layer value (using equation 4)

```
4. Evaluate the error at the neuron output
```
   $error_t = (x - t_{out})$
   Where, x is the desired output, $t_{out}$ is the actual output of the neuron

```
5. Apply gradient descent method to calculate error delta and adjust weights
from hidden layer to output layer
```
   $\delta_t = error_t * \frac{d(t_{out})}{d(t_{in})}$
   Where, $t_{in}$ is the weighted summation of all the hidden layer nodes
   $\delta_{w_j} \mathrel{+}= S_{j_{out}} \cdot \delta_t$

```
6. Backward propagate the output neuron error delta to calculate error delta at
hidden layer neuron and adjust weights from hidden layer to first layer
```
   $error_{s_j} = \delta_t \cdot w_j$
   $\delta_s = error_{s_j} * \frac{d(s_{j_{out}})}{d(s_{j_{in}})}$
   $\delta_{w_{ij}} \mathrel{+}= R_j \cdot \delta_s$

```
7. Go to step 4 for a certain number of iterations, or until the error is less
than a pre-specified value.
```

**Figure 6** Learning algorithm

In this implementation, a total of 36 synapses used to connect 8 input nodes to 4 hidden nodes and 1 output node. By training the network using learning algorithm and training dataset, 36 weights for these synapses found are shown in table 3. In the next section we will discuss the usage of these weights in DFM rule scoring.

**Table 3 Represent weightages for all the synapses**

|  | $S_1$ | $S_2$ | $S_3$ | $R_4$ |
|---|---|---|---|---|
| $T_{Out}$ | $w_1$ | $w_2$ | $w_3$ | $w_4$ |
| $R_1$ | $w_{11}$ | $w_{12}$ | $w_{13}$ | $w_{14}$ |
| $R_2$ | $w_{21}$ | $w_{22}$ | $w_{23}$ | $w_{24}$ |
| $R_3$ | $w_{31}$ | $w_{32}$ | $w_{33}$ | $w_{34}$ |
| $R_4$ | $w_{41}$ | $w_{42}$ | $w_{43}$ | $w_{44}$ |
| $R_5$ | $w_{51}$ | $w_{52}$ | $w_{53}$ | $w_{54}$ |
| $R_6$ | $w_{61}$ | $w_{62}$ | $w_{63}$ | $w_{64}$ |
| $R_7$ | $w_{71}$ | $w_{72}$ | $w_{73}$ | $w_{74}$ |
| $R_8$ | $w_{81}$ | $w_{82}$ | $w_{83}$ | $w_{84}$ |

In this flow synaptic weightages extracted from ANN is used. The flow uses input design to run DFM rule checking and generates DFM violations database. Similar to previous flow, this one also has context extraction module as to extract the respective layout context for all the DFM violations.

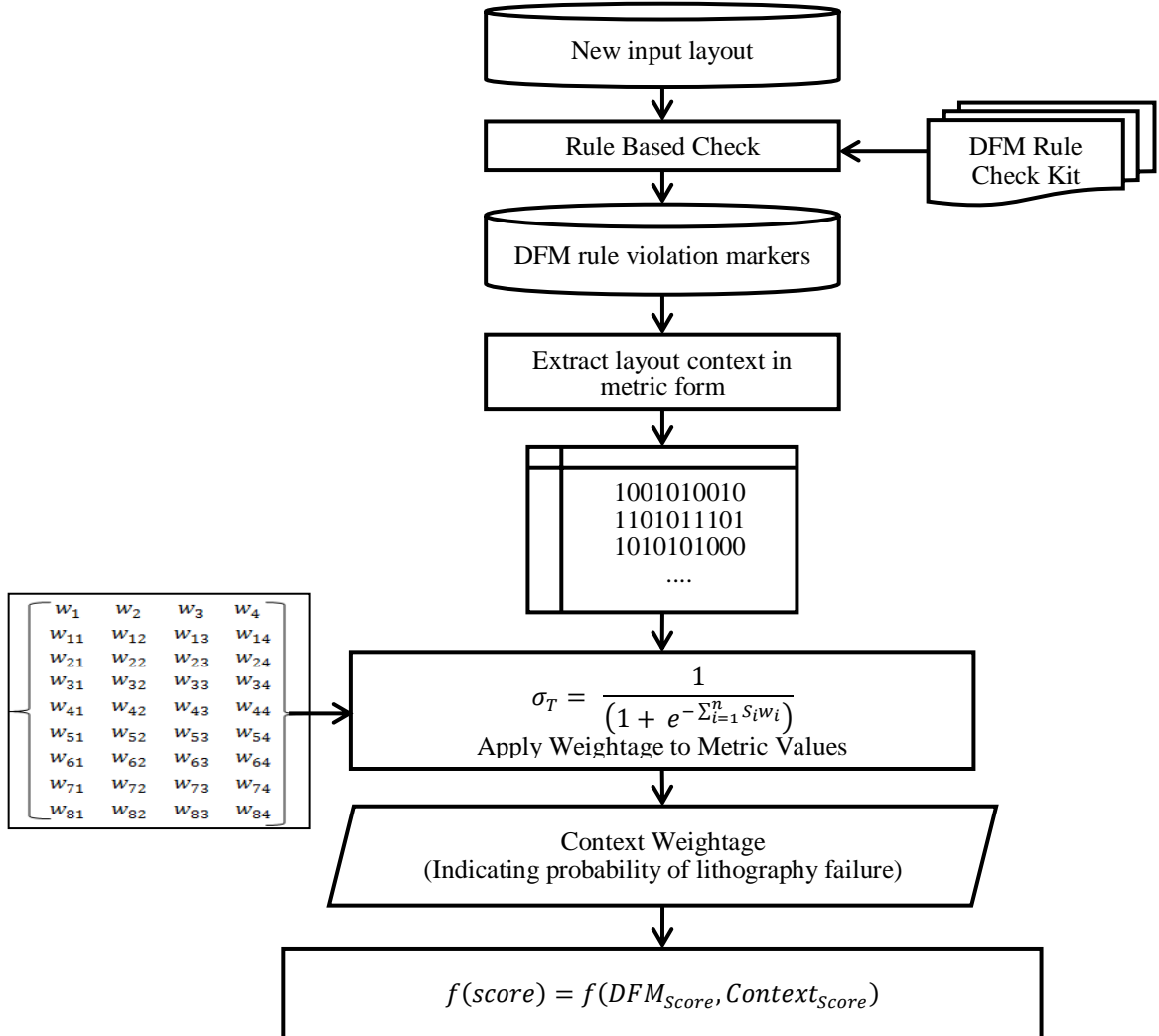

Figure 7 Context Arare DFM scoring flow

Weight is applied to the context values using equation 4. This provides the final weightage of the context which indicates probability of getting a lithography hotspot on that violation. Since sigmoid function is used in this approach, value varies from {0, 1}. Larger value signifies higher probability of failure while smaller values indicate lower probability of failure. This probability value is then converted in to context score using equation 6. By combining context score with original DFM score we get the new optimized DFM score, which can also be referred as Context Aware DFM Scores as shown in equation 7.

$$Context_{Score} = 1 - Context_{Weigtage} \quad \text{(Equation 6)}$$

$$Optmized_{Score} = f(DFM_{Score}, Context_{Score}) \quad \text{(Equation 7)}$$

From the earlier illustration, another score attached to the violation can be observed now. This score represent its context score. Applying equation 7, optimized Context aware DFM score is found. The violation with lower context score dominates over the existing DFM score and pulls down the final score close to zero, while the violation with higher context score improve the final score. These results now improve the correlation with lithography simulation results and become more lithography sensitive as shown in figure 8.

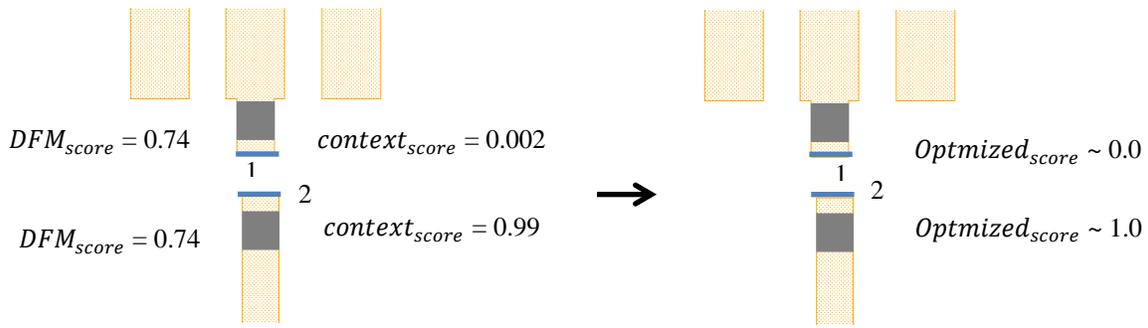

**Figure 8:** Example to demonstrate context aware scoring for two violations

### III. Implementation

To implement this flow, Mentor Calibre tool is primarily used with SVRF commands to perform DFM rule checking and score computation. Figure 9 shows sample code to check metal to via enclosure and compute score based on error width property followed by the DFM RDB command to save the marker in result database.

```
Via_err = RECTANGLE ENCLOSURE Via Metal SINGULAR OUTSIDE ALSO
GOOD EncValue  0.0 OPPOSITE EXTENDED EncValue EncValue 0.0 OPPOSITE
EXTENDED EncValue
Via_enc = ENC Via_err Metal < EncValue ABUT <90
Via_enc_prop = DFM PROPERTY Via_enc [DFM_Score = f(scoring_function)]
Metal_Via_Enc_err {
 DFM RDB M2_LE "Result.dfmdb" ALL CELLS NOPSEUDO
}
…..
DFM YS AUTOSTART Ext_Context_function
ContextExtraction
TVF FUNCTION Ext_Context_function [/*
proc ContextExtraction {} {
…
dfm::get_flat_geometries $topcell -layers layer_name -window $coord
dfm::add_property $iter "ContextScore" score_value -string
….
*/]
```

**Figure 9:** Sample code used in DFM checking, Context extraction and scoring annotation

From there on, Calibre Yield Server commands is used to read in the DFM result database and extract the DFM violation markers. These markers are used to evaluate layout context window and respective coordinates for each region by using "dfm::get_flat_gematries" command as shown in figure 9. The specifications of extracted geometries from each region later converted in metric form. This metric then used to analyze context by applying weightage module to evaluate the new context score. Further dfm::add_property command is used to annotate the new context scoring property to the corresponding DFM violation. In this way we can have new scoring calculated for each violation. In the next secession we will present context scoring for a test case and compare it with the conventional scoring results.

## IV. Results

To test the flow, a sample design having ~ 2000 DFM metal-via enclosure violation is used. Lithography simulation is performed to identify if there are any lithography hotspot present in the design. The simulation results show that there are total 8 lithography violation present in the design. When these lithography violation are overlaid with all the DFM rule violation then the lithography violation found to be under same bin along with other 91.63% of total violations. Figure 10 shows binning of DFM violations. This indicates the binning is not lithography aware as both, lithography frendly and lithography hotspot having the same score. Thismakes it hard to filter crtical violations from non critical ones.

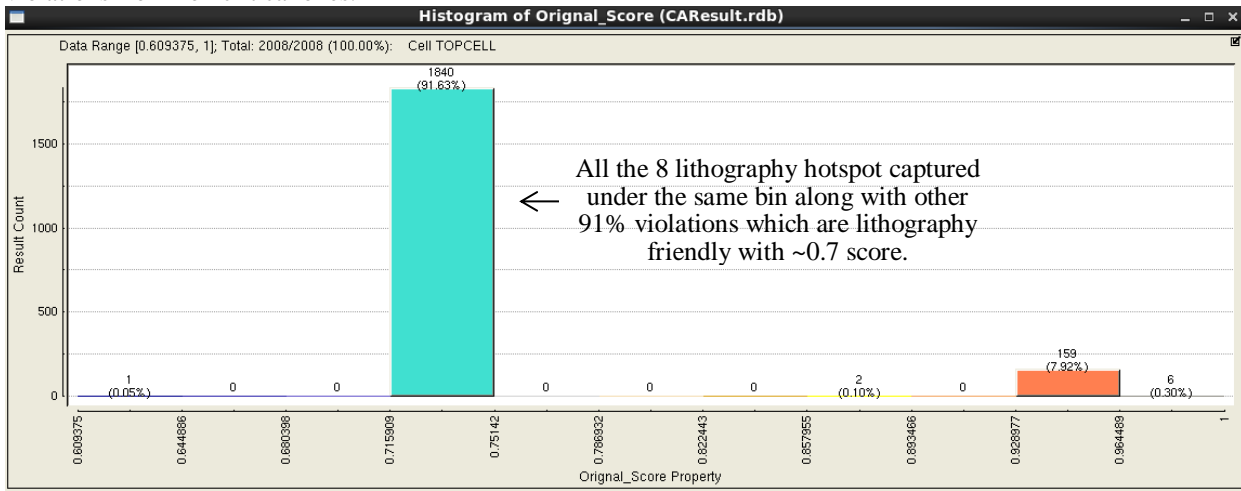

**Figure 10:** Showing Conventional DFM Score binning

With this proposed method, score can be optimized based on its context weigtage. Figure 11 shows the updated context aware scoring binning. All the lithography violations are clubbed together in smaller bin having only 2.79% of violations. These violations are more prone to be printibility issue and can be consider as potential weekpoints. On the other hand a big pool of violation been upgraded to higher scoring bin. This sort of binning helps in designing a focussed crtical violation by filtering based on score.

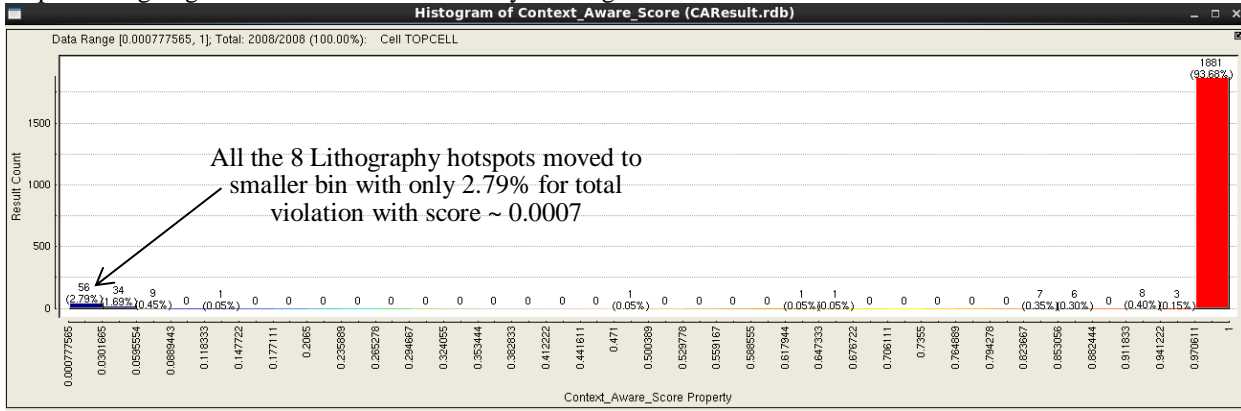

**Figure 11:** Context Aware Score binning**.**

## V. Conclusions

The Presented DFM scoring approach is crucial in terms of understanding design quality. Doing such scoring in a conventional way provides inaccurate results and misalignment with lithography simulation results. This also makes it hard to filter critical violations. The demonstrated context aware DFM rule scoring and analysis using machine learning technique resolve these issues and has many benefits. Specifically, the methodology is able to make rule-based checking approach more lithography aware, improves DFM rule score accuracy, helps designers to obtain critical violations, and help designers to improve design quality for better manufacturability.